\documentclass[sigconf]{acmart}

\usepackage{pdfpages}

\AtBeginDocument{%
  }

\copyrightyear{2026}
\acmYear{2026}
\setcopyright{cc}
\setcctype{by}
\acmConference[GamiFIN 2026]{10th Annual International GamiFIN Conference 2026}{March 23--27, 2026}{Saariselkä, Finland}
\acmBooktitle{10th Annual International GamiFIN Conference 2026 (GamiFIN 2026), March 23--27, 2026, Saariselkä, Finland}
\acmPrice{}
\acmDOI{10.1145/3789624.3789635}
\acmISBN{979-8-4007-2190-8/2026/03}

\begin{document}

\title
[Demonstrating SIMA-Play: Forest Management Decision-Making through Board Game and Digital Simulation]
{Demonstrating SIMA-Play: A Serious Game for Forest Management Decision-Making through Board Game and Digital Simulation}

\author{Arka Majhi}
\email{arka.majhi@tuni.fi}
\email{arka.majhi@uef.fi}
\orcid{https://orcid.org/0000-0002-5057-1878}
\affiliation{%
  \institution{Gamification Group\\Research Centre of Gameful Realities\\Tampere University\\University of Eastern Finland}
  \city{}
  \country{Finland}
}

\author{Daniel Fernández Galeote}
\email{daniel.fernandezgaleote@tuni.fi}
\orcid{https://orcid.org/0000-0002-5197-146X}
\author{Timo Nummenmaa}
\email{timo.nummenmaa@tuni.fi}
\orcid{https://orcid.org/0000-0002-9896-0338}
\author{Juho Hamari}
\email{juho.hamari@tuni.fi}
\orcid{https://orcid.org/0000-0002-6573-588X}
\affiliation{%
  \institution{Gamification Group\\Research Centre of Gameful Realities\\Tampere University}
  \city{Tampere}
  \country{Finland}
}

\author{Aaron Petty}
\email{aaron.petty@uef.fi}
\orcid{https://orcid.org/0009-0006-6595-1386}
\author{Jari Vauhkonen}
\email{jari.vauhkonen@uef.fi}
\orcid{https://orcid.org/0000-0002-2482-9644}
\author{Heli Peltola}
\email{heli.peltola@uef.fi}
\orcid{https://orcid.org/0000-0003-1384-9153}
\affiliation{%
  \institution{University of Eastern Finland}
  \city{Joensuu}
  \country{Finland}
}

\renewcommand{\shortauthors}{Majhi et al.}

\begin{abstract}
 Board games have shown promise as educational tools, but their use in engaging learners with the complex, long-term trade-offs of forest management remains strikingly underdeveloped. Addressing this gap, we investigate how forest growth simulation data can inform decision-making through information visualization and gameplay mechanics. We designed a serious game, SIMA-Play, that enables players to make informed forest management decisions under dynamic environmental and market conditions, simulating forest growth over time and comparing player performance across economic and sustainability outcomes. By using visualization to give players feedback on their choices, at the end of the game, it supports systems thinking and makes the trade-offs in forestry practices easier to understand and discuss. The study concludes with a research roadmap that outlines future experiments, longitudinal studies, and digital versions of SIMA-Play to assess its long-term effects on learning and engagement.

\end{abstract}

\begin{CCSXML}
<ccs2012>
   <concept>
       <concept_id>10003120.10003145.10003147.10010923</concept_id>
       <concept_desc>Human-centered computing~Information visualization</concept_desc>
       <concept_significance>500</concept_significance>
       </concept>
 </ccs2012>
\end{CCSXML}

\ccsdesc[500]{Human-centered computing~Information visualization}

\keywords{Forest, Ecosystems, Human-Forest Interaction, Forest Management, Sustainable Forestry, Sustainability, Sustainable Development Goal (SDG-15), Data-driven decision-making, SIMA model, Information Visualization, Data Visualization, Serious Game, Board Game, Game-based Learning}

\maketitle

\section{Introduction}

Forests play a crucial role in providing ecosystem services, including carbon sequestration, biodiversity, and clean water, in addition to supporting human livelihoods \cite{GlobalForestResourcesAssessmentFAO2020}. Under changing climatic conditions, the production levels of forest resources and ecosystem services face uncertainty \cite{Peltola2010}. Informed decision-making (DM) in forest management (FM) requires knowledge of forest ecosystem functions and attributes \cite{CharnleyAndPoe2007}. Incremental changes in forest structure and the long temporal periods in forestry increase uncertainties in FM. These changes are often induced by economic and policy shifts, as well as natural and artificial disturbances. To address these uncertainties, it is essential that FM stakeholders can make informed decisions and utilize tools that facilitate the understanding of management decisions and their potential outcomes. Such tools may help FM stakeholders in knowledge acquisition, learning by doing, and developing systems thinking over time \cite{Puettmann2009}. Traditional teaching methods often fail to demonstrate how forest ecosystems grow and respond, as well as the trade-offs associated with different management outcomes \cite{KrasnyAndMonroe2015}. To address this gap, researchers call for new teaching approaches that make the features of complex systems more communicable and easy to comprehend \cite{Waeber2023}.

Serious games, defined as games designed primarily for purposes beyond entertainment, such as education, training, or behavior change \cite{MichaelandChen2005} have shown promise in boosting engagement and knowledge retention \cite{Riopel2019}. Such games offer a safe space for experimentation, allowing players to explore complex models and promoting learning through interactive stories \cite{Gee2007}. In environmental contexts such as climate action, biodiversity preservation, and forest management, serious games have improved systems thinking, and risk awareness in DM\cite{Ahmadov2025}.

SIMA is a process-based forest ecosystem model developed in Finland to assess forest dynamics over long time horizons \cite{Kellomaki2008}. It integrates forest growth processes with FM actions such as thinning, regeneration, and harvesting. SIMA provides robust projections of biomass production, carbon balance, and economic yield under alternative management and climate scenarios. Due to the technical complexity, models like SIMA pose difficulty and are inaccessible to non-experts \cite{Lavalle2019}. In the context of gamified environments, simulation outputs can be translated into intuitive visual metaphors, interactive maps, and feedback systems that make abstract data more interpretable and meaningful for a broader audience \cite{HullmanAndDiakopoulos2011}. Our study presents the design of an FM game based on the SIMA model, aiming to enhance engagement, systems thinking, and DM. 

We designed SIMA-Play, an analogue serious game which is also an Eurogame format board game \cite{Eurogames2012}. Eurogames (also called german-style board games) are focused on indirect player competition(e.g., through shared resources rather than direct attacks), minimal randomness (e.g., dice rolls), resource optimization, and long-term planning as seen in \textit{Catan} (resource trading), \textit{Agricola} (farming resources) and \textit{Terraforming Mars} (planetary development). These features also closely relate to FM, helping players to engage in complex decision processes balancing trade-offs among ecological, economic, and social objectives, rather than relying on luck or player elimination. Eurogame structure provides a framework for simulating realistic management scenarios and supporting players in systems thinking. SIMA-Play can be classified as a strategy-oriented resource management Eurogame, focusing on sequential DM, multi-criteria optimization, and adaptive responses to dynamic environmental risks.

SIMA-Play imitates the DM processes in FM in Finland, a country with a high percentage of privately owned forests and complex governance structures involving multiple actors \cite{Hujala2015}. Through this research, we aim to contribute to the growing field of game-based FM education, offering a model for integrating forest simulation and information visualization to address the challenges of informed DM in forestry under uncertainty.

\section{Background}

Education plays a crucial role by increasing people’s knowledge of forest ecosystems and encouraging sustainable choices \cite{KrasnyAndMonroe2015}. However, traditional methods often fall short of communicating the complex and dynamic aspects of FM. Thus, there is growing interest in more interactive and engaging teaching approaches.

Interactive data visualization (IDV) enhances users’ ability to quickly and accurately interpret complex information. In the context of games, visual representations of data can serve both aesthetic and functional purposes, guiding player behavior, signaling feedback, and conveying consequences. IDV such as real-time maps \cite{Thomas2022}, dashboards\cite{Helbig2022}, and infographics \cite{Becklas2023} have been used effectively in games like \textit{SimCity} \cite{ArnoldSimCity2019} and \textit{Civilization} \cite{VidalCivilizations2020} to model environmental systems and policy impacts \cite{Pottinger2023}. In educational games, IDV facilitates exploratory learning by enabling users to manipulate variables and observe the resulting outcomes. Research in Human-Computer Interaction suggests that combining narratives with IDV \cite{HullmanAndDiakopoulos2011} can deepen engagement and comprehension \cite{SegelAndHeer2010}. When applied to forest ecosystems, this means visualizing not only tree cover or species distribution but also more abstract concepts such as ecosystem services \cite{Powley2023}, carbon sequestration \cite{Lindrup2023}, or policy trade-offs \cite{Shavazipour2022,Vergarechea2023}.

Simulation modeling has been used extensively in FM to predict growth, yield, and the impact of various interventions. The SIMA model is one such framework that supports DM in forestry. SIMA emphasizes iterative learning and scenario-based forecasting, making it suitable for educational contexts. SIMA is a gap-type forest ecosystem simulator \cite{Kellomaki2008} designed to model regional predictions of forest growth by simulating the carbon and nitrogen cycles with changing climate conditions. Tree growth is modeled based on individual trees reacting to their environment. This allows for the simulation of forest dynamics under various management and climate scenarios.

By integrating different scenarios with SIMA-modelled outputs within game environments, complex ecological data can be translated into interactive formats, making them more accessible and engaging for users. This approach can enhance understanding of forest ecosystems and support DM processes in FM. However, simulation tools often remain inaccessible to non-expert users due to their technical complexity. Translating simulation outputs into easy-to-understand IDV is a key challenge and opportunity for Human Computer Interaction and game design \cite{Buono2020, Su2021}. By embedding SIMA-based scenarios within game environments with the help of IDV, we can make abstract data tangible, relatable, and actionable. For example, forest growth data can be visualized through dynamic landscapes that respond to player decisions, while biodiversity indices can inform win/loss conditions or resource availability in the game.

The convergence of serious game, simulation modeling, and IDV offers a promising path forward for teaching/exploring FM. While each of these domains has its own body of research, few studies integrate them into cohesive, user-centered interventions. Our work builds on this nascent interdisciplinary space, designing a system that aims to make scientific models accessible, forest growth data meaningful, and sustainability challenges playable.

%\section{Methods}

\section{Overview of SIMA-Play}

SIMA-Play is a Eurogame-style, turn-based experiential learning tool designed for promoting strategic DM and systems thinking in forest management. In this section, we discuss the rules and the underlying mechanics of the game. The design of the gameboard, play objects, and rulebook are appended in the Appendix. \ref{reference}

\subsection{Game Board and Initial Setup}

The physical game board has 40 equal squares, each representing one hectare of forested land. Players choose a particular set of colored sticky notes and stick them on the squares to indicate their active management and maintain visual differentiation during play. Each participant starts the game with ten parcels (10 hectares).

\begin{figure}[!htp]
    \centering
    \includegraphics[width=\linewidth]{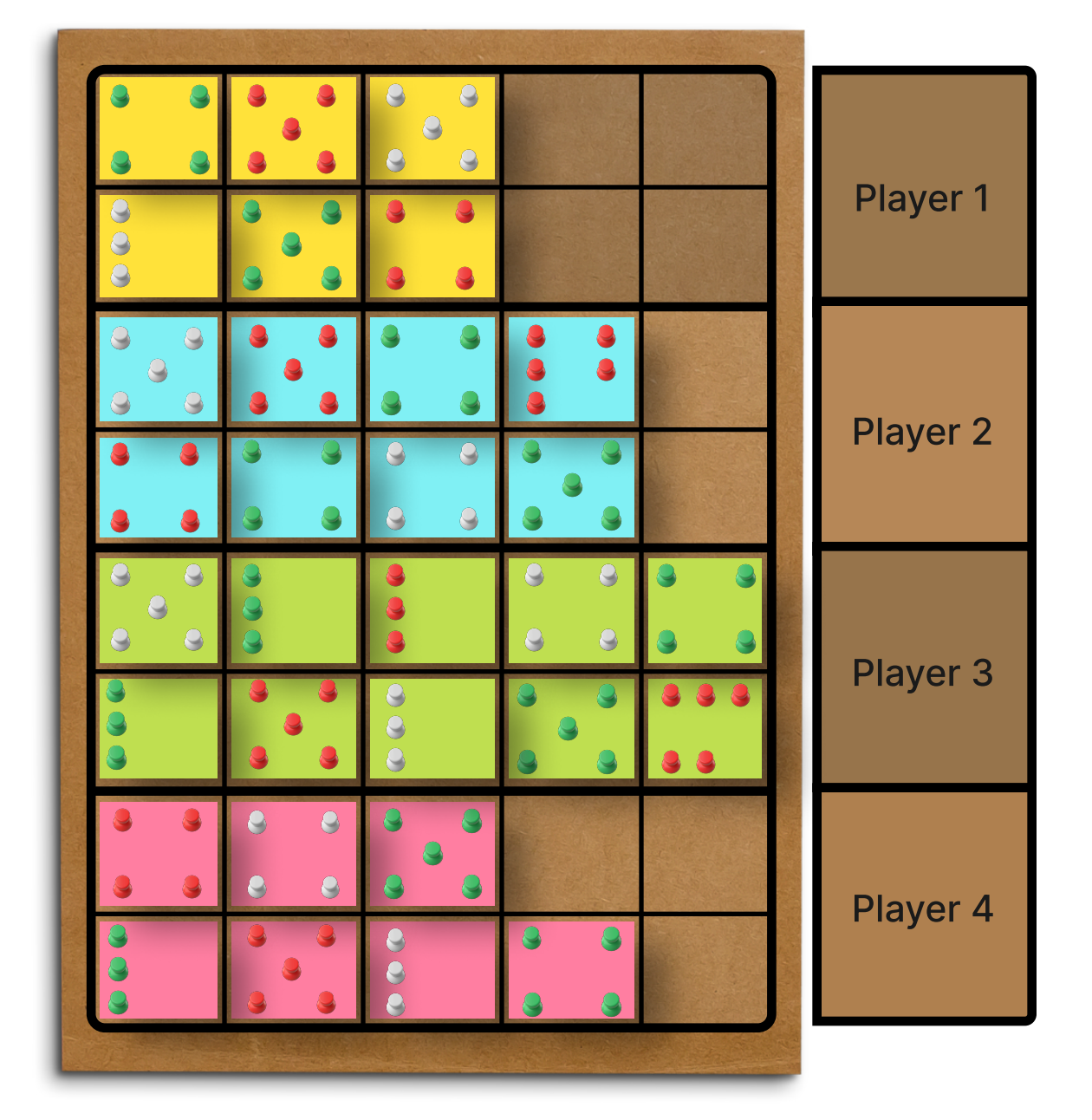}
    \caption{Game Board of SIMA-Play}
    \label{fig:placeholder}
\end{figure}

At the start, every player receives an initial allocation of €8,000 in game tokens, representing available capital for investment and operations. This currency enables players to conduct a range of financial transactions, including purchasing or leasing land parcels and investing in FM activities. Effective fund management mirrors the economic constraints and opportunities inherent in real forestry enterprises.

\subsection{Species Selection and Planting Phase (Year 0)}

During the initial planting phase (Year 0), players are allotted ten hectares of forest land. We choose the common Nordic species, Scots pine (\textit{Pinus sylvestris}), Norway spruce (\textit{Picea abies}), and Silver birch (\textit{Betula pendula}), for their distinct characteristics and resilience profiles. To make forest composition tangible, we use colored thumbtacks to represent trees. We use red pins for Scots pine, green pins for Norway spruce, and white pins for Silver birch. Each pin symbolises 400 trees, with a maximum planting density of 2,000 trees per hectare, corresponding to 5 pins per parcel. Players are required to select a single species per parcel, although they are free to diversify species across their parcels of land to improve resilience and economic stability.

The cost of planting is €1,000 per hectare, creating an early strategic choice, whether to invest all available capital in planting or to reserve funds for future management and unforeseen events. Players can also lease out parcels to other players for the full 60-year game duration, receiving immediate cash in return. The introduction of elements of negotiation and cooperation can support interaction and competition.

\subsection{Economic Progression and Yield Phases}

Each game round is a 60-year virtual timespan, which represents four major decision points. Year 0 - Planting Phase, Year 30 – First Commercial Thinning, Year 45 – Second Commercial Thinning, and Year 60 – Final Felling/Harvest. These four stages mimic FM cycles, each requiring different FM actions, investment decisions, and adaptive responses to risks/uncertainty. Between each round, players assess their positions, calculate revenues and expenses, and adjust future strategies accordingly.

The game simulates economic returns and operational costs through three successive phases of timber harvesting. In the first commercial thinning (Year 30), 1,000 trees per hectare are harvested, yielding 50 m³ of pulpwood per hectare, resulting in a revenue of €1,000 (€20/m³). Similarly, in the second commercial thinning (Year 45): 600 trees per hectare are harvested, yielding 50 m³ pulpwood and 50 m³ sawwood per hectare, resulting in a revenue of €1,000 (pulpwood) and €2,500 (sawwood €50/m³). In the final felling/harvest (Year 60), 400 trees per hectare are harvested, yielding 50 m³ pulpwood and 150 m³ sawwood per hectare, resulting in a revenue of €1,000 (pulpwood) and €9,000 (sawwood).

\subsection{Uncertainty and Risk Simulation}

In order to simulate real-world unpredictability, a deck of 'multi-risk cards' is introduced. These cards represent both natural disturbances (such as mammal grazing, bark beetle infestations, storms, or fungal decay) and economic shocks (such as fluctuations in timber prices). After each DM phase, players draw one card from the multi-risk deck, introducing an element of chance, risk, and adaptive management.

For example, in the first turn, drawing a Mammal Damage card can lead to the loss of approximately 40\% of young pine and birch saplings (1–2 meters in height), through grazing by moose and reindeer. In gameplay terms, this event requires players to remove two pins (40\%) out of five from all affected parcels containing these species.

To manage these risks, players can purchase insurance, which costs €500 if bought between 0–30 years, €1,000 between 31–45 years, and €2,000 between 46–60 years. Once purchased, the insurance is valid for 1 parcel of land till final harvest. This provides coverage against major disturbances such as bark beetle attacks (affecting spruce), storm damage, and fungal root rot. Insurance decisions are voluntary, leading players to strategically compare risk mitigation and cost control. If a multi-risk card indicates a bark beetle infestation, the sawwood value from spruce is downgraded to the price of pulpwood unless insurance coverage is purchased beforehand. Likewise, market fluctuation cards can increase or decrease timber prices by €10 per m³, affecting all players equally. These mechanics encourage players to continually adapt FM strategies, taking into account both ecological risk and economic volatility across extended temporal horizons.

\subsection{Engagement, Learning, and Motivation}

The dynamic interplay of strategy, uncertainty, and social interaction intends to create fun and engagement in SIMA-Play. Players try to strike a balance between interdependent goals, short-term profits, and long-term sustainability. The multirisk cards introduce unpredictability to the gameplay, as players react to unforeseen events, leading to increased excitement and engagement. Financial constraints, insurance choices, and competitive bargains encourage strategic thinking and collaboration. From a pedagogical perspective, SIMA-Play promotes experiential learning by linking DM to tangible outcomes. Players not only learn the mechanics of FM economics but also internalize the ecological consequences of FM strategies under uncertainty. The visual and tactile elements, such as colored pins and sticky notes, that represent land parcels further reinforce cognitive engagement, transforming abstract forestry concepts into interactive and memorable experiences. SIMA-Play aims to create a balance between educational depth and playful engagement by motivating players through curiosity, challenge, and the satisfaction of mastering complex systems within a shared, competitive environment.

\subsection{Comparison through Simulation and Interactive Data Visualization}

Following the completion of each gameplay session, the decision logs from all players are input into the SIMA forest growth simulation model. This integration enables the translation of gameplay decisions into quantitative estimates of biophysical and economic outcomes, including tree biomass carbon, total soil carbon, ecosystem carbon, wood products carbon, timber, deadwood, soil water, and net present value.

\begin{figure}[!htp]
    \centering
    \includegraphics[width=\linewidth]{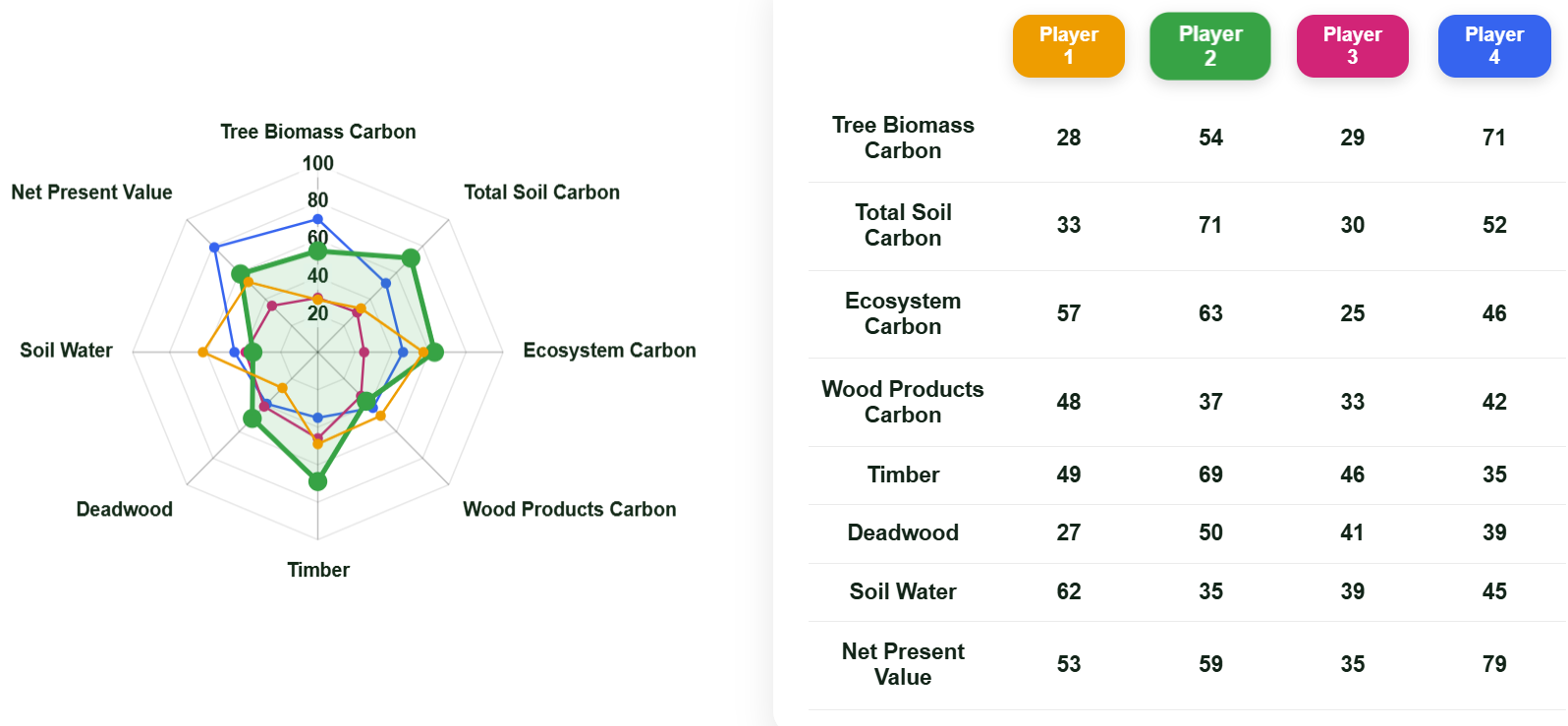}
    \caption{Mockup of Interactive Data Visualization of Post-Gameplay Scores Depicting Player Decision Outcomes from SIMA-Play (Scaled 1–100)}
    \label{fig:placeholder}
\end{figure}

Costs, revenues, and net profits are computed based on the simulated volumes and associated market conditions. Each group reviews its SIMA-generated output visualizations, providing immediate feedback and allowing players to assess the long-term implications of their management strategies.

Further analysis of SIMA simulation results evaluates DM patterns, more detailed accounts of economic profitability, and ecosystem service outcomes across different management strategies. Descriptive and inferential statistics can identify trends, significant differences, and correlations between player decisions (e.g., species choice, risk tolerance, investment behavior) and resulting ecosystem metrics.

IDV interface supports players in exploring how their choices influenced various outcome indicators. This enables the identification of distinct player behaviors, such as those prioritizing economic gain, non-timber ecosystem services, or balanced multi-objective management. Through SIMA-Play, players can directly observe the implications of their virtual management decisions, bridging experiential learning with analytical understanding.

\section{Conclusion}
The study shows how SIMA-Play, an analogue hybrid serious game can support game-based learning and enable systems thinking. The game design exemplifies how concepts such as the cycle of growth, risk management, or the conflict between immediate gain and long-term sustainability in FM can be integrated into a game design without oversimplifying underlying scientific models. The post-game use of IDV supports learning by allowing players to compare strategies, reflect on outcomes, and connect DM with quantitative ecosystem outcomes. The study also highlights its potential as a learning tool, and as a communication tool, in FM to compare outcomes due to uncertainty.

\section{Future Work and Research Directions}

This study constitutes the initial phase of research focused on designing and demonstrating SIMA-Play, a board game prototype that harnesses computational outputs from the SIMA model simulator to advance FM education. Building on the prototype's foundational integration of analogue gameplay with simulation-driven feedback, subsequent phases will rigorously evaluate its pedagogical efficacy through targeted empirical investigations.

Longitudinal investigations, spanning 3–6 months with multi-session protocols, can assess the retention of learning outcomes through repeated gameplay, delayed post-tests (e.g., at 1, 3, and 6 months), and semi-structured follow-up interviews. These can probe retention of systems-level understanding—particularly adaptive responses to uncertainty and transfer to analogous real-world contexts, such as scenario planning for climate-impacted forests. Mixed-methods analysis will integrate quantitative metrics with qualitative thematic coding to elucidate mechanisms of long-term knowledge consolidation and behavioral change.

Concurrently, we plan to explore digital enhancements (e.g., tablet-based ecological dashboards and overlays) to deliver continuous, context-sensitive feedback; these investigations will compare such modalities against SIMA outputs to determine their relative impact on engagement, reflection, and collaboration in the long term. Iterative co-design workshops with end-users will refine these features, prioritizing scalability for low-resource FM training environments.

The demonstration is expected to stimulate community discourse on best practices for integrating complex forest growth modeling into tangible interfaces and to offer a replicable roadmap for researchers seeking to hybridize embodied and computational learning environments in forest management.

\begin{acks}
The authors acknowledge the use of OpenAI’s ChatGPT (version: GPT-4o) for generating the artworks used in the game cards and rulebook illustrations. The tool was used under the authors’ supervision, and all content, design decisions, and interpretations are the sole responsibility of the authors, in accordance with the ACM Policy on Authorship and the Use of Generative AI.

This research was supported by the UNITE flagship (2025-2026, decision 359172 (UEF), 359173 (TAU)) and the European Union – NextGenerationEU instrument, through the Multirisk project [grant number 353263] funded by the Research Council of Finland.
\end{acks}

\bibliographystyle{ACM-Reference-Format}
\bibliography{sample-base}

\appendix
%\chapter{Appendix}


\section*{Supplementary material}

# A  Appendix: How to Play - Rule Book for SIMA-Play

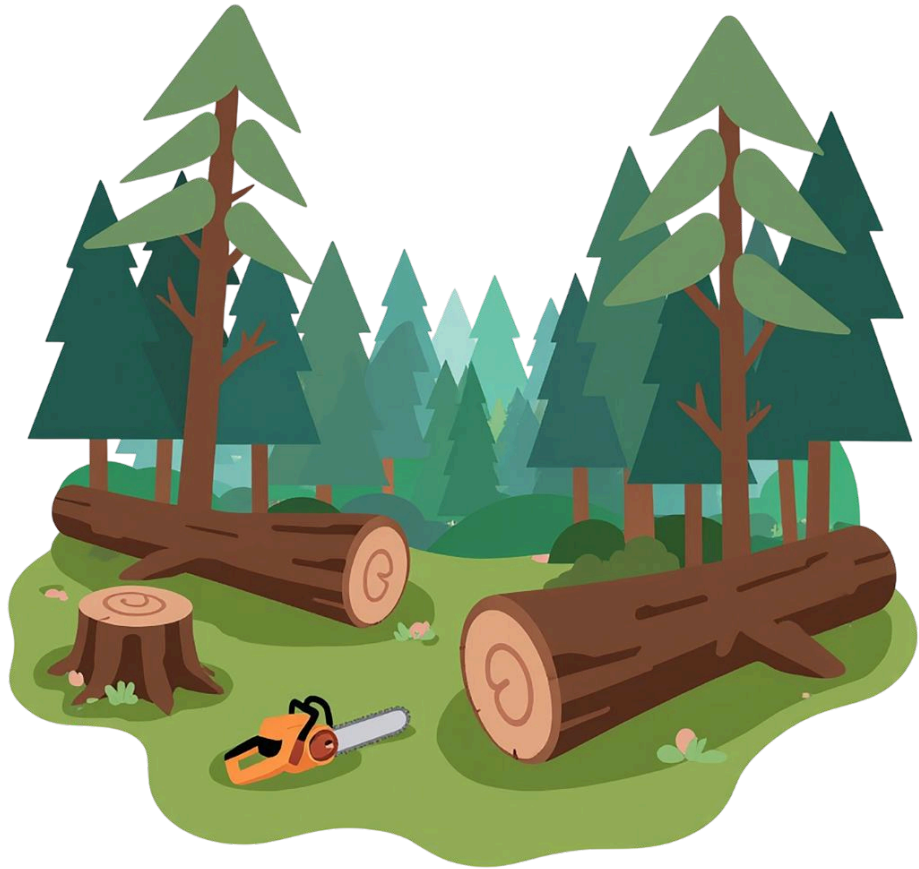

# SIMA Play

## Learning Forest Management through Play

# GAME COMPONENTS

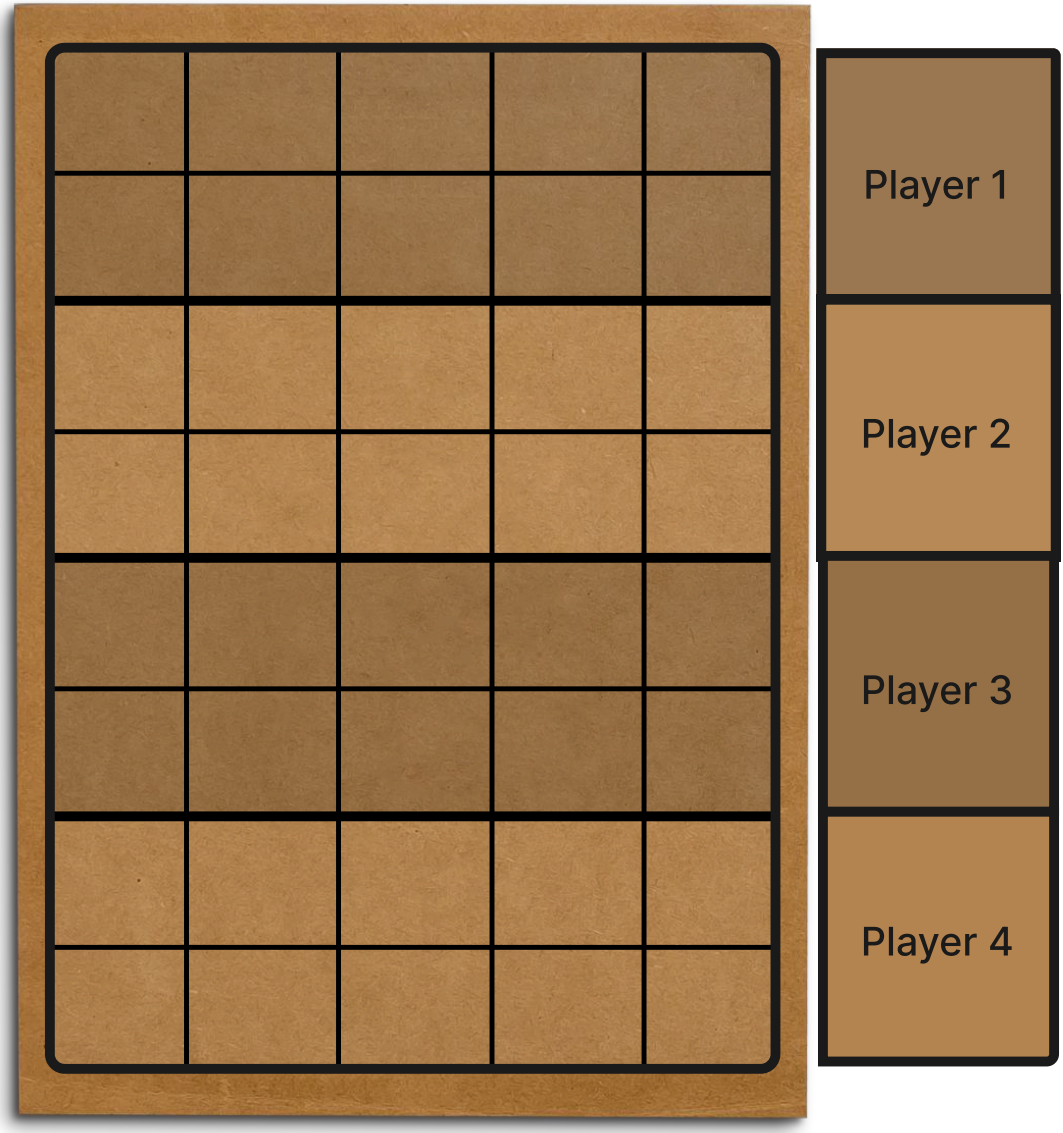

**Gameboard**

The board has 40 squares representing parcels of land. Each parcel is equal to 1 hectare. In a 4-player game, each player gets 10 parcels of land to manage

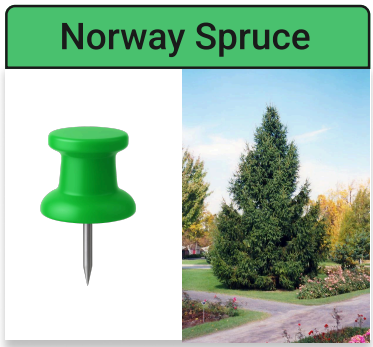 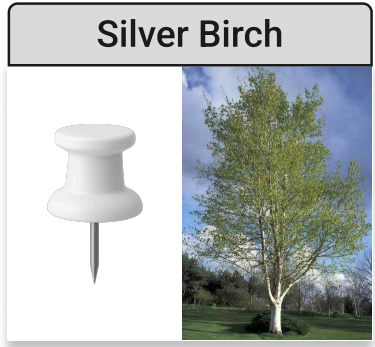 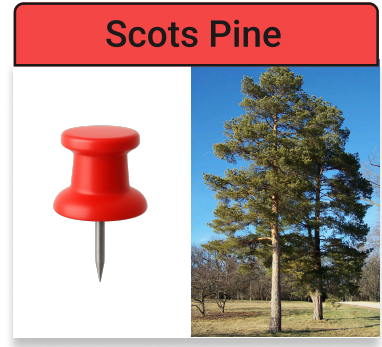

## Plant Species

Colored pushpins represent different tree species: Red for Scots Pine, Green for Norway Spruce & White for Silver Birch. One pushpin represents 400 trees

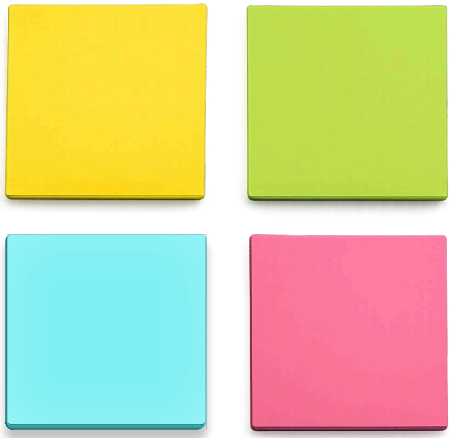

Each player is assigned 10 sticky notes of their chosen color, which are placed on land parcels to represent active management

## Sticky Notes

In-game currency tokens have three values: 5000 for large tokens, 500 for medium tokens, and 50 for small tokens.

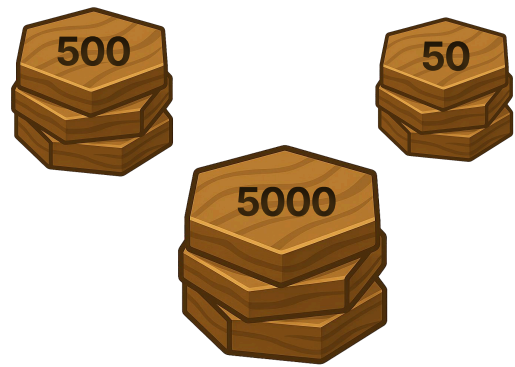

## In-Game Currency Tokens
8

Page 2-13

# Risk Cards

**PRICE DECREASE**

When drawn, this card creates fluctuations in timber prices.

Decreasing market prices of timber affects the growth strategy of plants

**MAMMAL GRAZING**

The damage caused by storms affects all the three species of trees.

Remove two pins from each parcel of land that contains any of the three species

The effect on the damage can be prevented using Remedy card or Insurance

**BARK BEETLE**

This infestation affects only the *Norway Spruce* trees

Remove two pins from each parcel of land that contains any of the three species

The effect of damage can be prevented using Remedy card or Insurance

**ROOT DISEASE**

The root rot in this card is caused by fungus called *Heterobasidion*

A fall in saw wood prices relative to pulpwood 60€ — 25€

**PRICE INCREASE**

When drawn, this card creates fluctuations in timber prices.

Increasing market prices of timber affects the growth strategy of plants

**STORM DAMAGE**

Storm damage affects all three tree species

Remove three pins from each parcel of land that contains any of the three species

The effect of damage can be prevented using Remedy card or Insurance

9

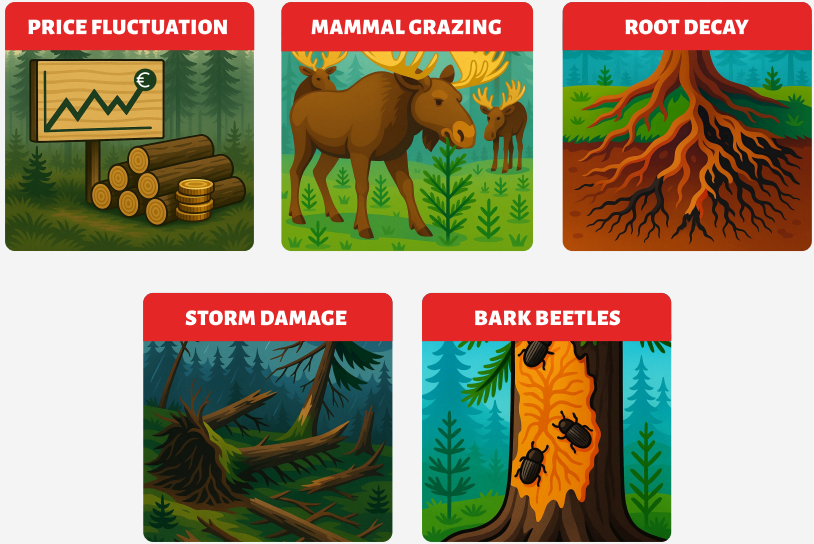

**Remedy Card**

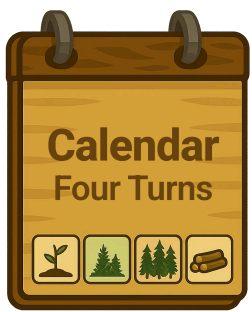

**In-Game Calendar**

The in-game calendar keeps a record of the four decision-making turns for all players

10

Page 4-13

## YEAR 0
### Planting Phase

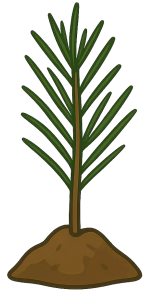

1st turn in calendar

## YEAR 30
### First Commercial Thinning

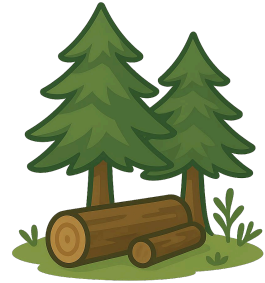

2nd turn in calendar

## YEAR 45
### Second Commercial Thinning

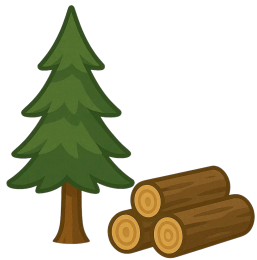

3rd turn in calendar

## YEAR 60
### Final Felling/Harvest

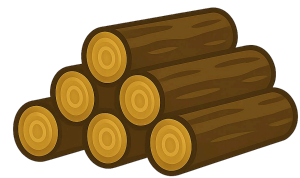

4th turn in calendar

# INSURANCE CARD

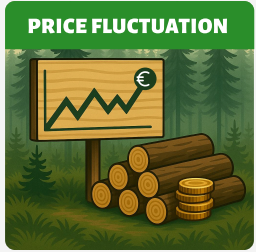
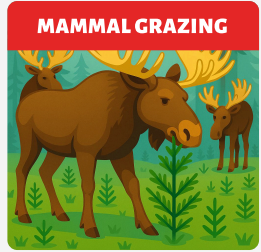
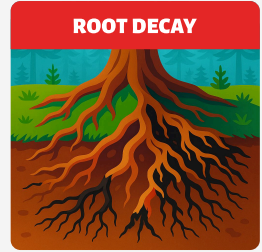
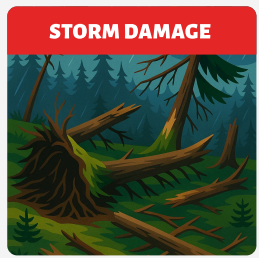
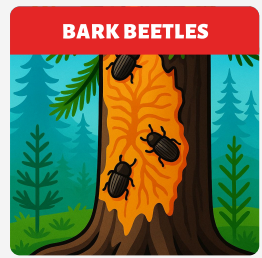

## Cost

The cost of the card depends on the time of purchase:
0-30 years - 500 €
31-45 years - 1000 €
46-60 years - 2000 €
Once purchased, the insurance is valid for one parcel of land until the final harvest

## Function

This card minimizes or prevents risks caused by risk cards

Insurance Card

# Anatomy of a Risk Card

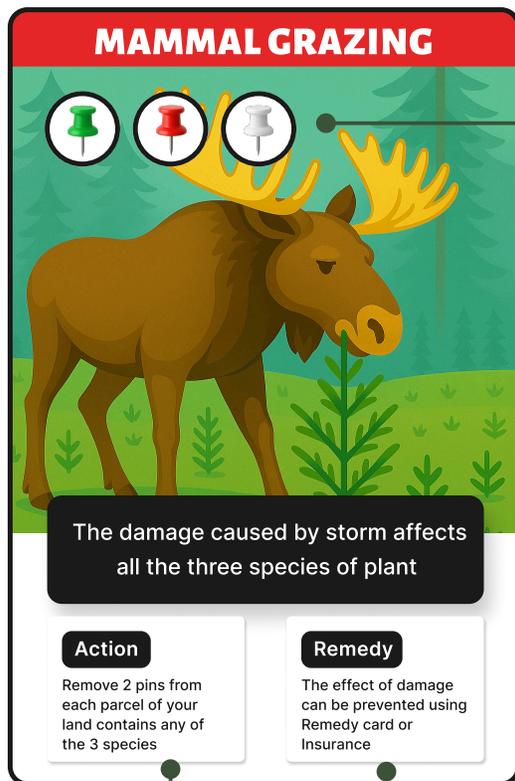

Icons represent the tree species affected by the risk on the card

The action section represents the steps players take when the card is drawn

Remedy Cards neutralize the effect of Risk Cards

# Setting-Up

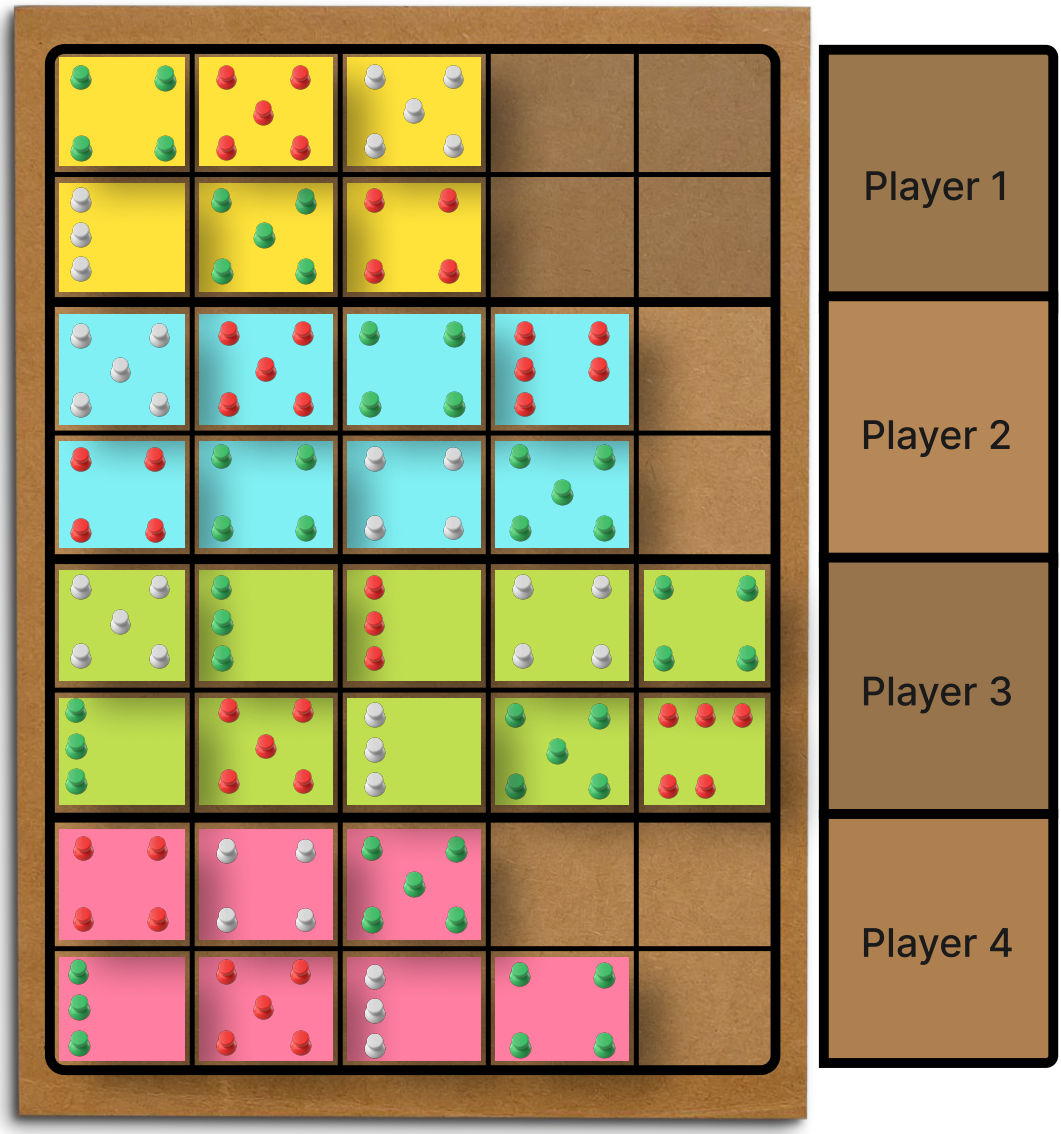

# Setting-Up

- Each player receives sticky notes of their chosen color, which are placed on the parcels of land they choose to manage. On the sticky notes, players place the colored pushpins representing the tree species they select in the first turn

- Players can only select one type of tree species out of the three that they can plant on a parcel of land. Players can choose one of these densities to plant trees in each parcel of land:

| 1 Pin = 400 trees | 2 Pins = 800 trees | 3 Pins = 1200 trees |

| 4 Pins = 1800 trees | 5 Pins = 2000 trees |

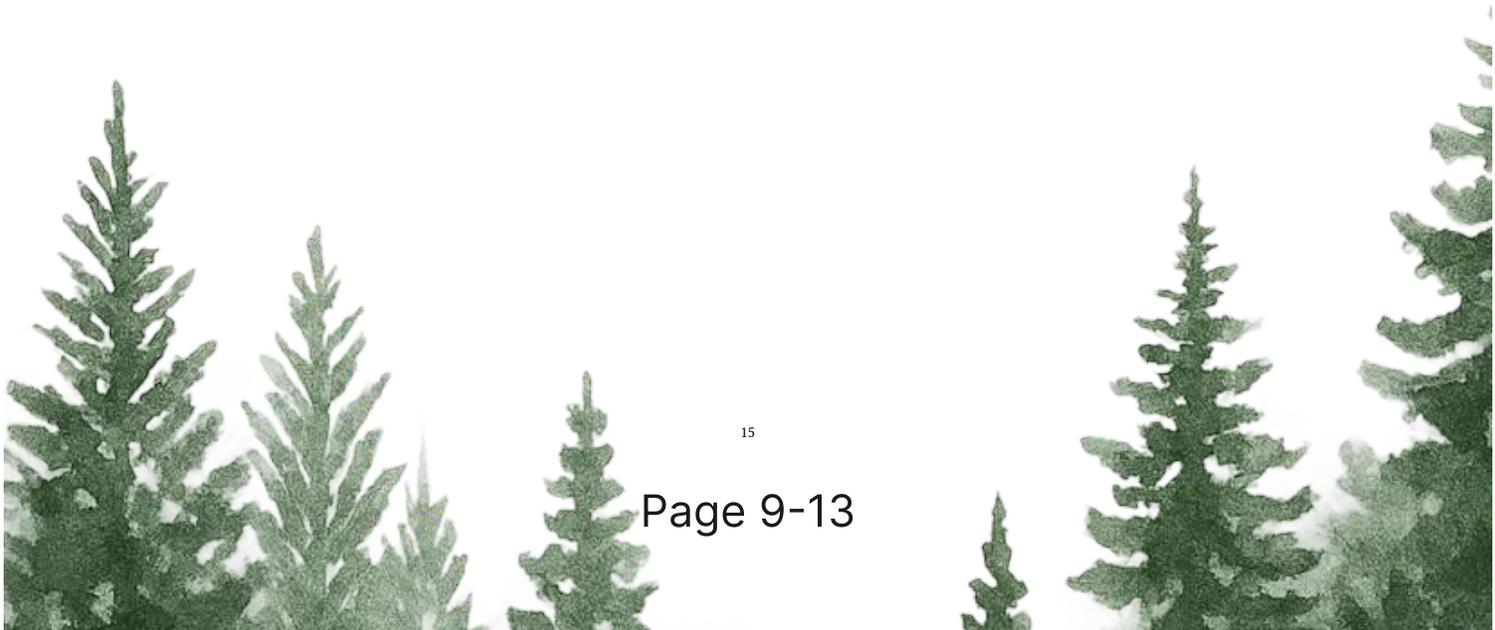

15

Page 9-13

# Conditions

- All players grow and manage forests on each parcel of land over a period of 60 years

- Players record their decisions in the form of notes on paper

- At every turn, players review forest growth and draw a card from the risk card deck.

- All recorded decisions are entered into the SIMA simulation model at the end of the game to calculate scores of each player

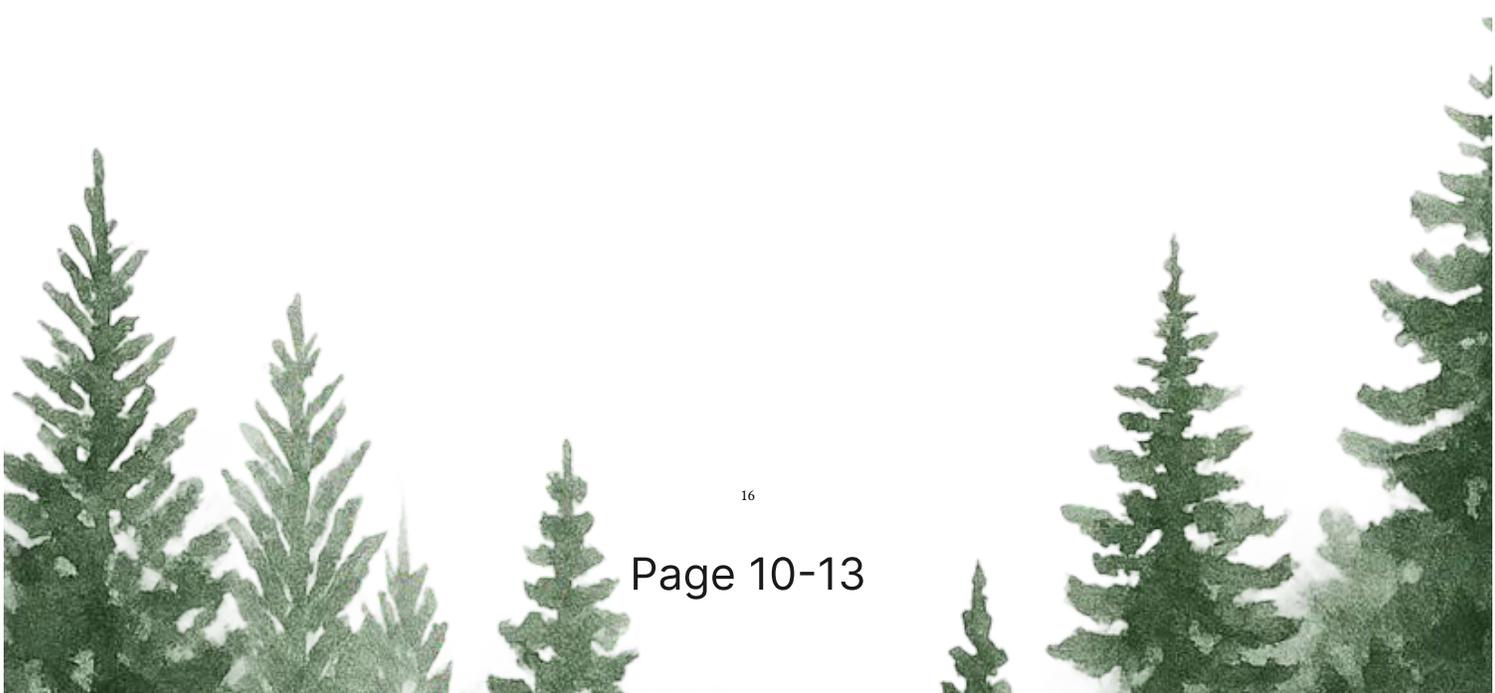

16

Page 10-13

# Rules

- All the players will grow and manage the forest on each parcel of land on a course of 60 years

- Throughout the game, four turns are played. Year 0 - Planting Phase, Year 30 – First Commercial Thinning, Year 45 – Second Commercial Thinning, and Year 60 – Final Felling/Harvest

- Each player is initially given 10 parcels of land and decides how many parcels they want to manage

- In each turn, players perform various forest management activities such as thinning to ease overcrowding, pruning for better timber quality and final harvests

- At first turn, players are allowed to purchase, sell or lease their parcels of land among other players

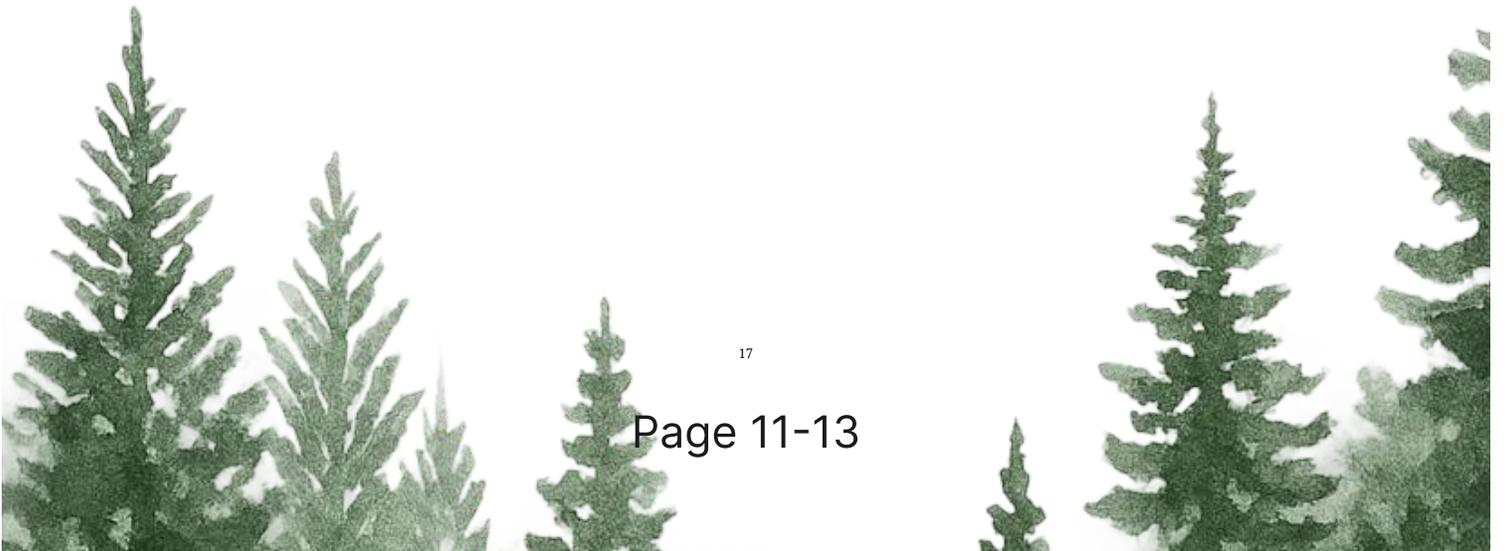

17

Page 11-13

# Rules

- After all players complete their turns in a row, the last player flips the calendar to the next year to start the next turn

- The event that is mentioned in the card such as bark beetle damage outbreaks, storm damage and artificial price fluctuations will change the outcome and the payoffs

- At the end of the game, the players are scored by SIMA model on the basis of the decisions taken by players during the gameplay

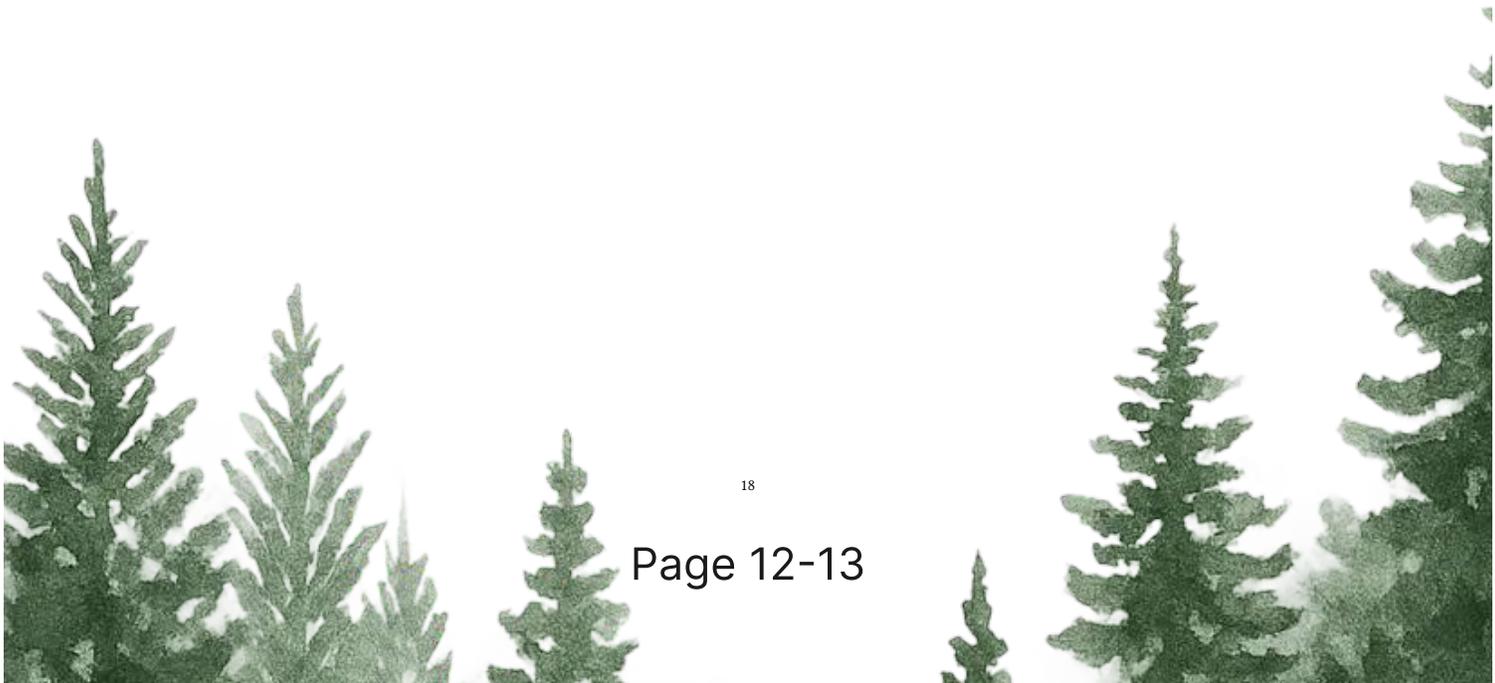

# Win/Lose

- There is no win-or-lose criterion

- Each player is scored on different parameters like timber volume, total carbon, soil water, etc.

- Player scores are represented through interactive radar charts

- Each player can compare their scores with those of other players

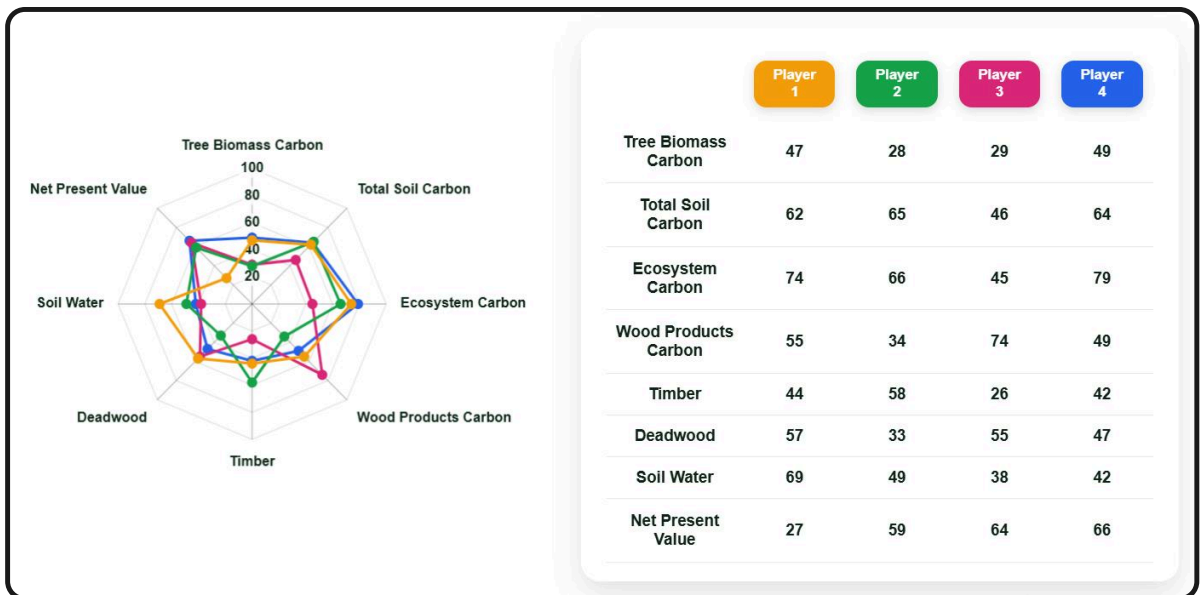



\end{document}